\documentclass[11pt]{article}
\usepackage{moriond,epsfig}

\def\Journal#1#2#3#4{{#1} {\bf #2}, #3 (#4)}

\def\NIM{\em Nucl. Instrum. Methods}

\def\be{\begin{equation}}
\def\ee{\end{equation}}
\def\bea{\begin{eqnarray}}
\def\eea{\end{eqnarray}}

\begin{document}
\vspace*{4cm}
\title{THE LHCB COMMISSIONING}

\author{ S. DE CAPUA (for the LHCb collaboration) }

\address{European Organization for Nuclear Research,\\
CERN, CH-1211 Gen{\`e}ve 23, Switzerland}

\maketitle\abstracts{
The LHCb experiment is dedicated to precision measurements of CP violation and rare decays of B
hadrons at the Large Hadron Collider (LHC) at CERN. The LHCb installation has been finished in 
spring 2008 and an intensive testing and commissioning of the system has been started.
An overview and the results from our commissioning activities are described.}

\section{Introduction}\label{sec:intro}
LHCb~\cite{magn1,ReopTDR} is an experiment dedicated to heavy flavour physics at the LHC. 
Its primary goal is to look for indirect evidence of new physics in CP violation and rare decays
of beauty and charm hadrons. With the large $\rm b\overline{b}$ production cross section of $\sim
500\,\mu$b expected at an energy  of 14 TeV,  the LHC will be the most copious source of B mesons 
in the world. With a modest luminosity of $2 \times 10^{32}$~$\rm cm^{-2} s^{-1}$ for LHCb, 
$10^{12}$ $\rm b \overline{b}$ pairs would be produced in $10^7$ s, corresponding to one year of 
data taking. The luminosity for the LHCb experiment can be tuned by changing the beam focus at its
interaction point independently from the other interaction points. This will allow LHCb to maintain 
the optimal luminosity for the experiment for many years from the LHC start-up. 
%
%
%
%
%

\section{Detector overview}\label{sec:detector}
\begin{figure}[ht]
\begin{center}
\psfig{figure=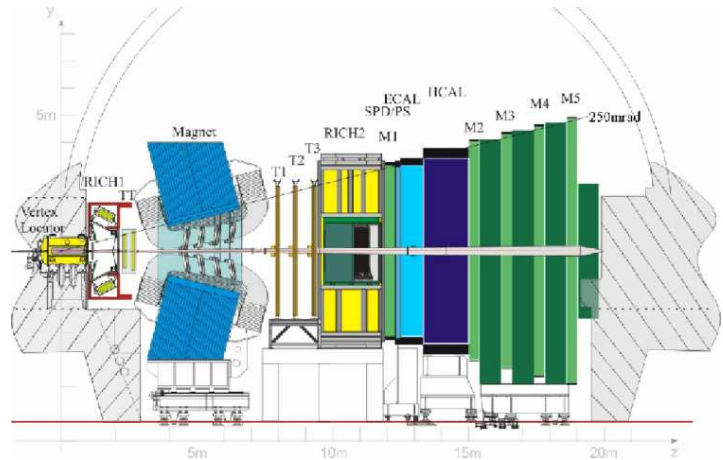,height=60mm}
\caption{View of the LHCb detector.}
\label{fig:lhcb_layout}
\end{center}
\end{figure}

LHCb is a single-arm spectrometer with a forward angular coverage from approximately 10 mrad to 300 
(250) mrad in the bending (non-bending) plane. The choice of the detector geometry is justified by the fact 
that at high energies both the b- and $\rm \overline{b}$-hadrons are predominantly produced in the same 
forward or backward cone. The layout of the LHCb spectrometer is shown in figure~\ref{fig:lhcb_layout}.

The LHCb detector is made of various detector subsystems. The vertex locator system (including a pile-up 
veto counter), called VELO, is a silicon detector which surrounds the interaction point and aims at precise
measurements of the radial and angular position of the tracks. The spectrometer magnet is a warm dipole, 
providing an integrated field of 4 Tm. The tracking system is made up of a Trigger Tracker 
(a silicon microstrip detector, called TT) in front of the magnet, and three tracking stations behind 
the magnet, made of silicon microstrips for the inner parts (IT) and of Kapton/Al straws for the outer 
parts (OT).
To achieve excellent $\pi$--K separation in the momentum range from 2 to 100 GeV/c, the LHCb detector is
equipped with two Ring Imaging Cherenkov counters (RICH1 and RICH2), which use Aerogel, C${_4}$F$_{10}$ 
and CF${_4}$ as radiators. The calorimeter system is composed of a Scintillator Pad Detector and 
Preshower (SPD/PS), an electromagnetic (shashlik type) calorimeter (ECAL) and a hadronic (Fe and scintillator 
tiles) calorimeter (HCAL). Finally, the muon detection system providing $\mu$ identification and contributing 
to the Level-0 trigger of the experiment, is composed of MWPC's (except in the highest rate region, where 
triple-GEM's are used).
%

\subsection{The LHCb trigger}\label{sec:trigger}
As mentioned in sec~\ref{sec:intro}, the LHCb experiment plans to operate at an average luminosity which 
is much lower than the maximum design luminosity of the LHC, reducing the radiation damage to the 
detectors and electronics. 
At LHC, the crossing frequency is 40~MHz, which has to be reduced by the trigger to about 
2~kHz, at which rate the events are written to storage for further offline analysis.
This reduction is achieved in two trigger levels \cite{TDRTrigger,L0NIM2}: the Level-0 (L0) and the High Level 
Trigger (HLT). 

The L0 trigger is implemented using custom made electronics, operating synchronously with 
the 40~MHz bunch crossing frequency. Its purpose is to reduce the LHC beam crossing rate of 40~MHz to 
the rate of $1$~MHz with which the entire detector can be read out. 
The Level-0 trigger attempts to reconstruct (i) the highest $E_{\rm T}$ hadron, electron and photon clusters
in the calorimeters and (ii) the two highest $p_{\rm T}$ muons in the muon chambers.
%

The HLT is instead executed asynchronously on a processor farm, using commercially available equipment,
and it makes use of the full event data in order to be able to reduce the event rate from 1~MHz down to 2~kHz. 
%
Since the HLT is fully implemented in software, it is very flexible and will evolve with the knowledge 
of the first real data and the physics priorities of the experiment. 

\section{LHCb commissioning}\label{sec:commissioning}
The LHCb commissioning started in 2007. The main purposes were initially to test and calibrate each detector 
subsystem individually and finally run the LHCb detector as a whole.

In the first phase, the safety issues have been checked and hardware operations controls and monitoring 
have been tested. Then, calibration pulses have been used to test the response of the hardware, which
allowed to check control cables, data cables and trigger signals. Calibration pulses have been also used
to check the channel mapping and to spot any anomalous electronic channels (dead or noisy). 
From the data produced by each sub-detector, the initial settings of time and spatial alignment 
had been set to reasonable values.

Once all the detector subsystems have been commissioned, the system was exercised as a whole. 
The LHCb experiment has been read out to a maximum hardware trigger rate of 100~kHz
due to the limited capacity of the network and the event filter farm. The designed hardware trigger
rate of 1~MHz will be reached in 2009. Data storage at 2 kHz was already exercised in 2008. 

\subsection{Commissioning with cosmic rays events}\label{subsec:cosmic}
\begin{figure}[ht]
\begin{center}
\psfig{figure=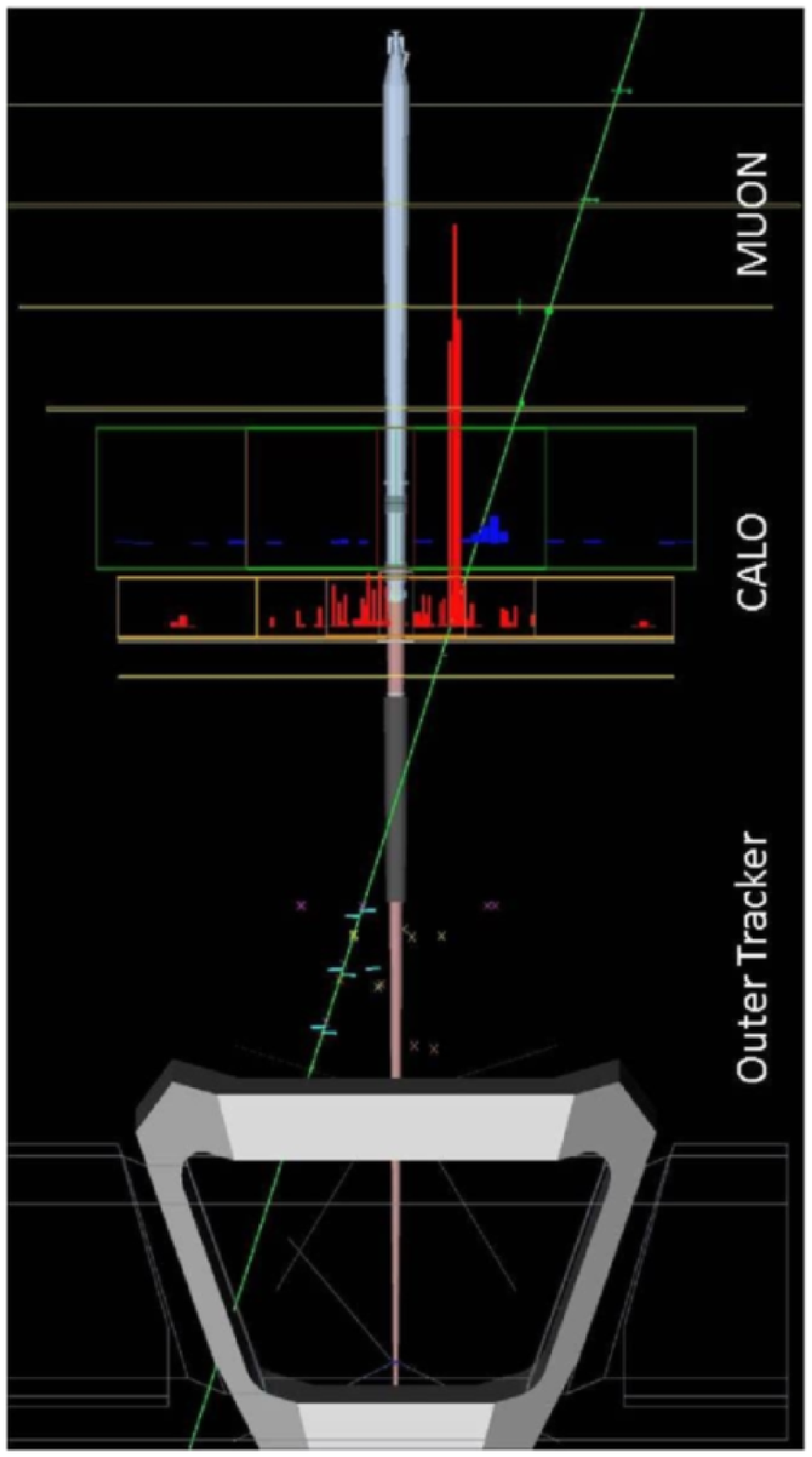,height=60mm}~~~~~~~~
\psfig{figure=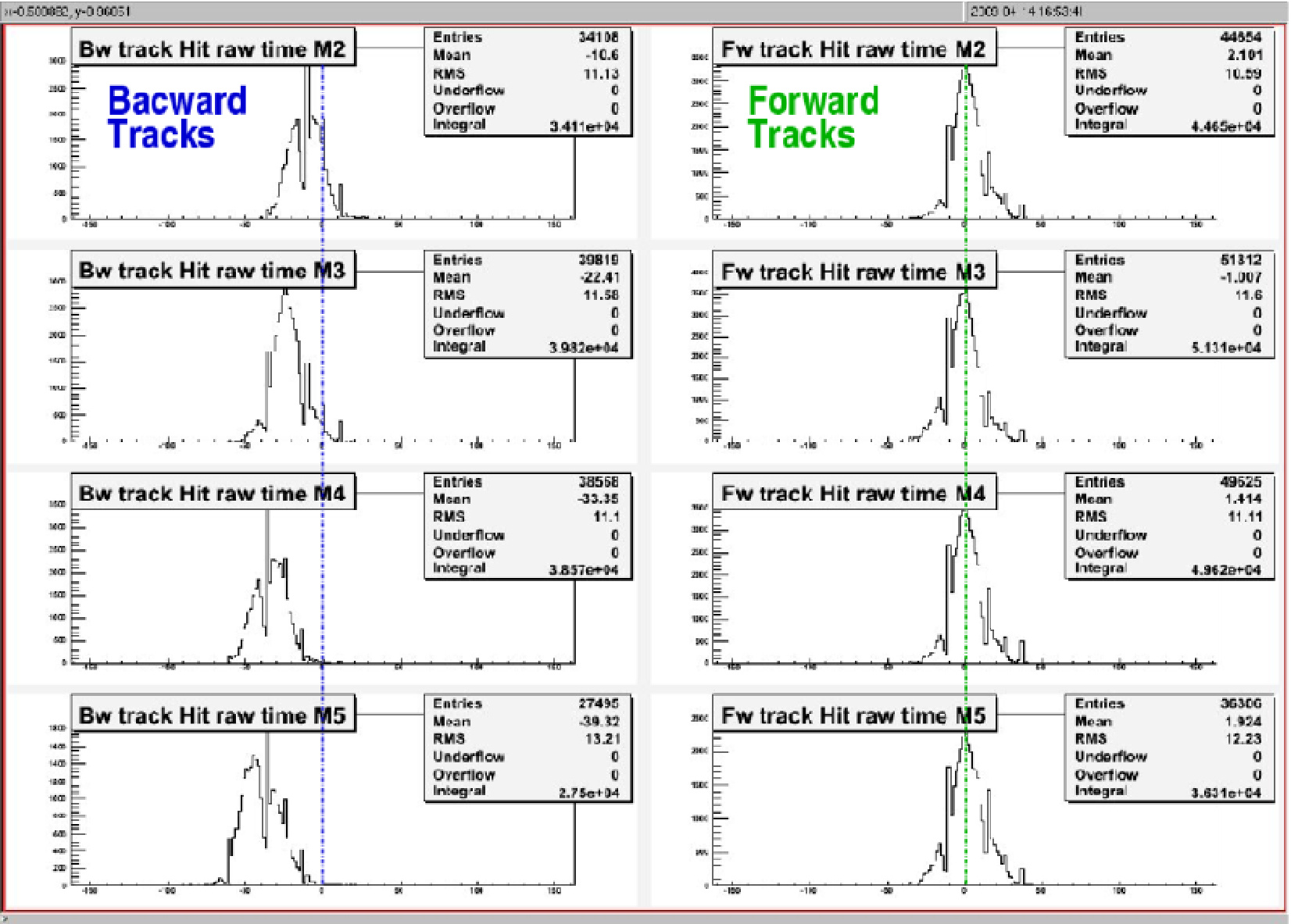,height=60mm}
\caption{(Left) Event display of a cosmic event reconstructed by the OT, Calo and Muon systems. 
(Right) Time distribution of four muon stations (M2-M5) with backward and forward tracks.}
\label{fig:cosmic}
\end{center}
\end{figure}

Although the configuration of the LHCb experiment is not well suited for cosmic runs (the rate of tracks 
within $\pm$250 mrad from horizontal is well below 1~Hz), more the a million useful cosmic events
have been recorded. A typical cosmic event is shown in figure~\ref{fig:cosmic} (left).

In order to acquire cosmic events, the ECAL/HCAL and MUON trigger has been slightly relaxed: 
(i) thresholds on $p_{\rm T}$ and $E_{\rm T}$ have been lowered in order to be sensitive to minimum
ionizing particles, (ii) the coincidence of only two muon stations has been required and (iii) tracks
have not been constrained to point to any vertex. 

Cosmic data have been extremely useful to commission the trigger (the same logic as for real data has been used)
and to achieve a spatial alignment and a coarse time alignment of one detector or a few neighbors in which 
more vertical cosmic tracks can be used. In figure~\ref{fig:cosmic} (right), the time distribution of four
muon stations (M2-M5) with respect to the trigger time is shown: the trigger time was given 
by the SPD and optimized for forward tracks. 

However, not all the detectors could be aligned with this method: in particular, the Trigger Tracker and 
the Vertex Locator are too small and too far from the detectors providing the trigger, and they could not 
participate in the cosmic runs. 

\subsection{Commissioning with beam induced events}\label{subsec:beam}
In summer 2008, the LHC machine group performed two sector tests: one beam was injected from the 
SPS into the LHC to test the new injection lines and their optics. The beam was dumped in a stopper (TED) 
about 340~m downstream the LHCb experiment. The fluence was about 10~particles/cm$^2$. 
%
In a second step, the beam was stopped 50~m from the LHCb detector in a collimator/beam stopper
(TDI) just after the kicker injection magnet: in this case the fluence was 100 times higher.
Finally, the beam was injected and passed through the LHCb experiment. During the last two steps, the 
subdetectors have been kept off most of the time to avoid damage.

\begin{figure}[ht]
\begin{center}
\psfig{figure=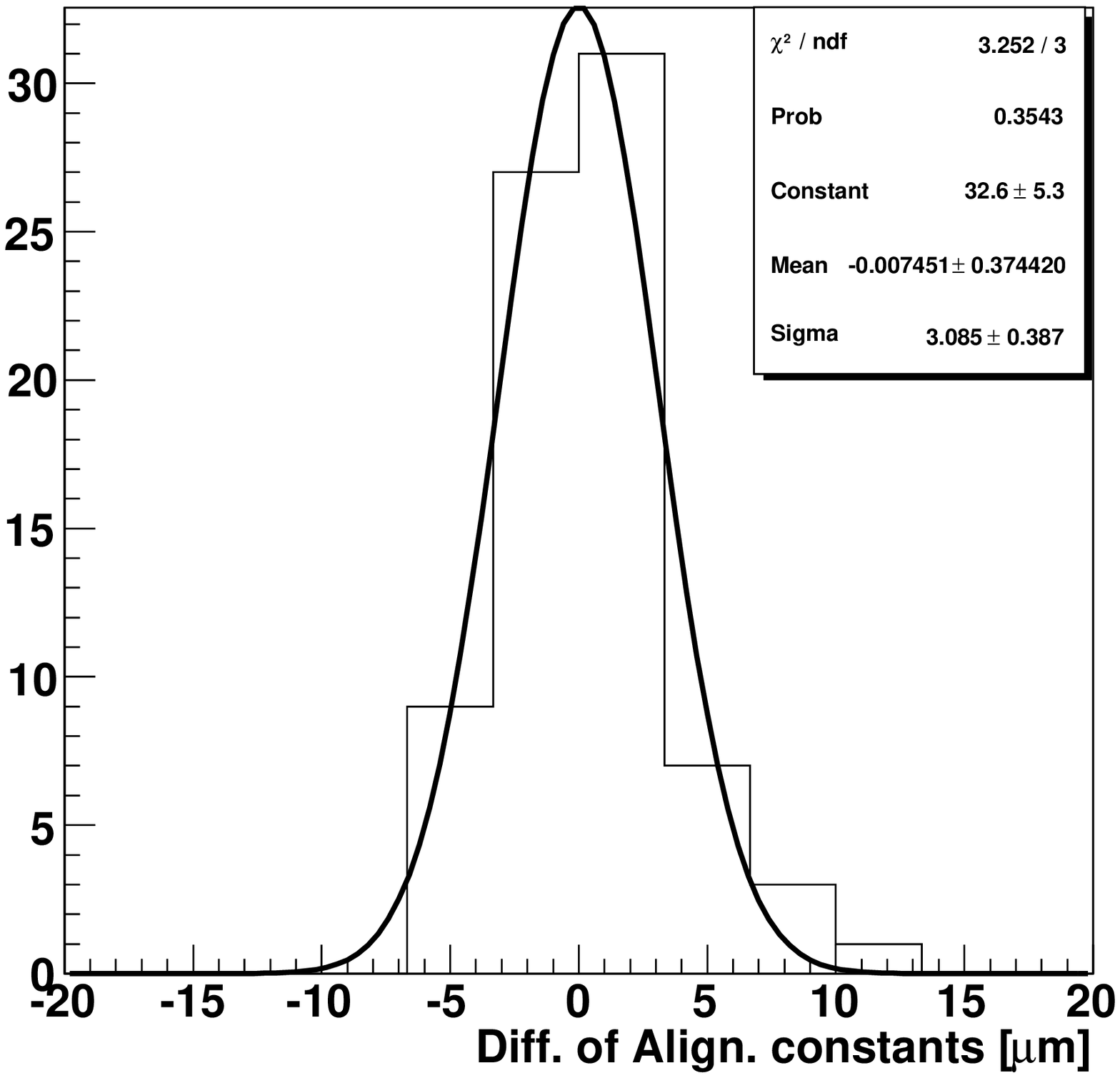,height=55mm}~~~~~~~~
\psfig{figure=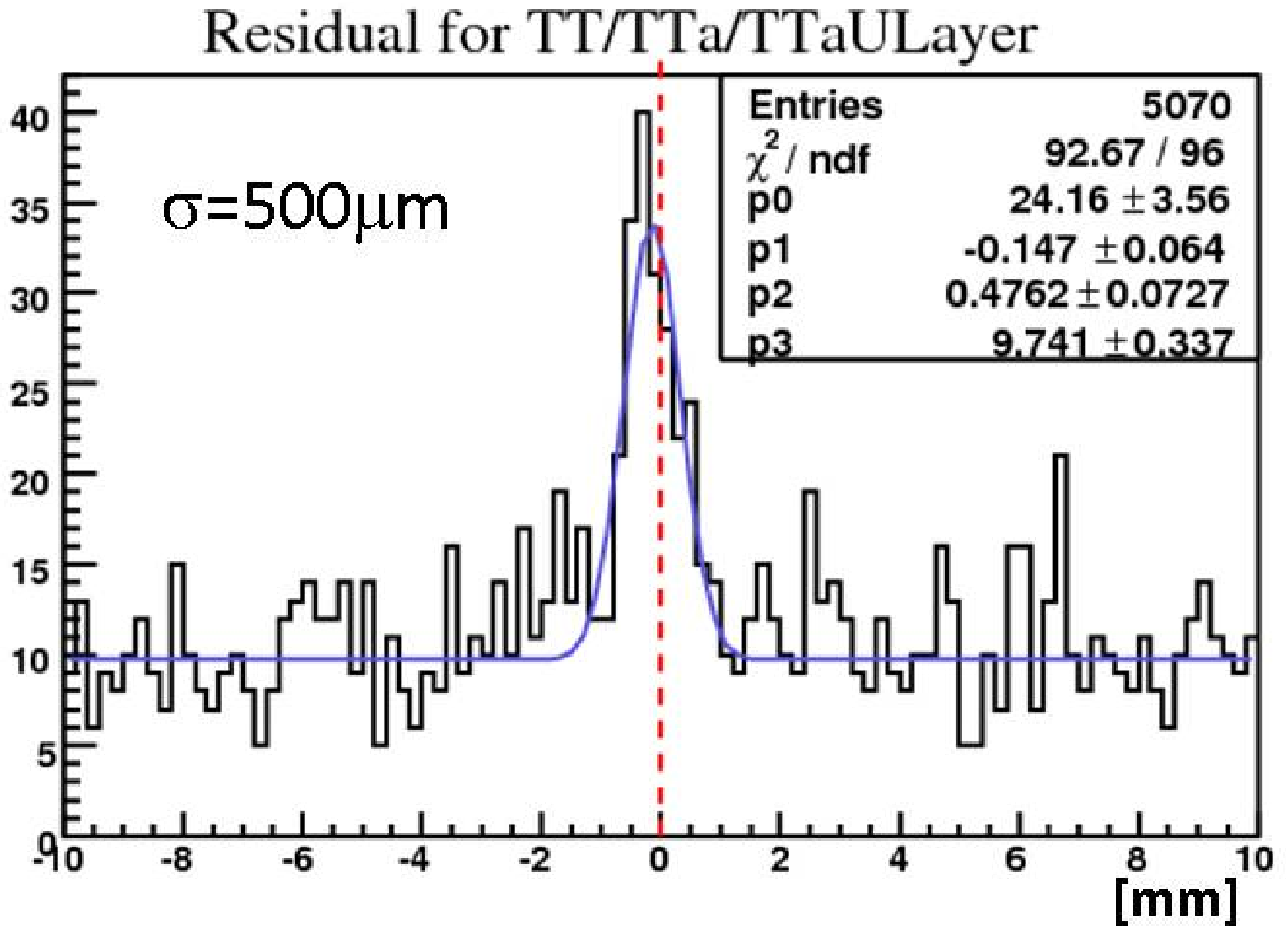,height=50mm}
\caption{VELO (left) and TT (right) spatial alignment with beam induced events. Note that the large background 
in the right plot is due to the high fluence of particles.}
\label{fig:SpaceAlign}
\end{center}
\end{figure}

Beam induced data have been used to perform a more precise time and spatial alignment, especially for those 
detectors which could not make use of cosmic events. Figure~\ref{fig:SpaceAlign} (left) shows the distribution 
of the difference between the alignment constants computed with two different data samples: the VELO spatial 
alignment resolution is 5~$\mu$m for X(Y) translations and 200~$\mu$rad for Z rotations~\cite{parkes}. 
Once reconstructed in the Vertex Locator, tracks have been then extrapolated to the Trigger Tracker and the hit 
residuals have been computed: the resolution is about 500~$\mu$m and the result is shown in 
figure~\ref{fig:SpaceAlign} (right). 

\begin{figure}[ht]
\begin{center}
\psfig{figure=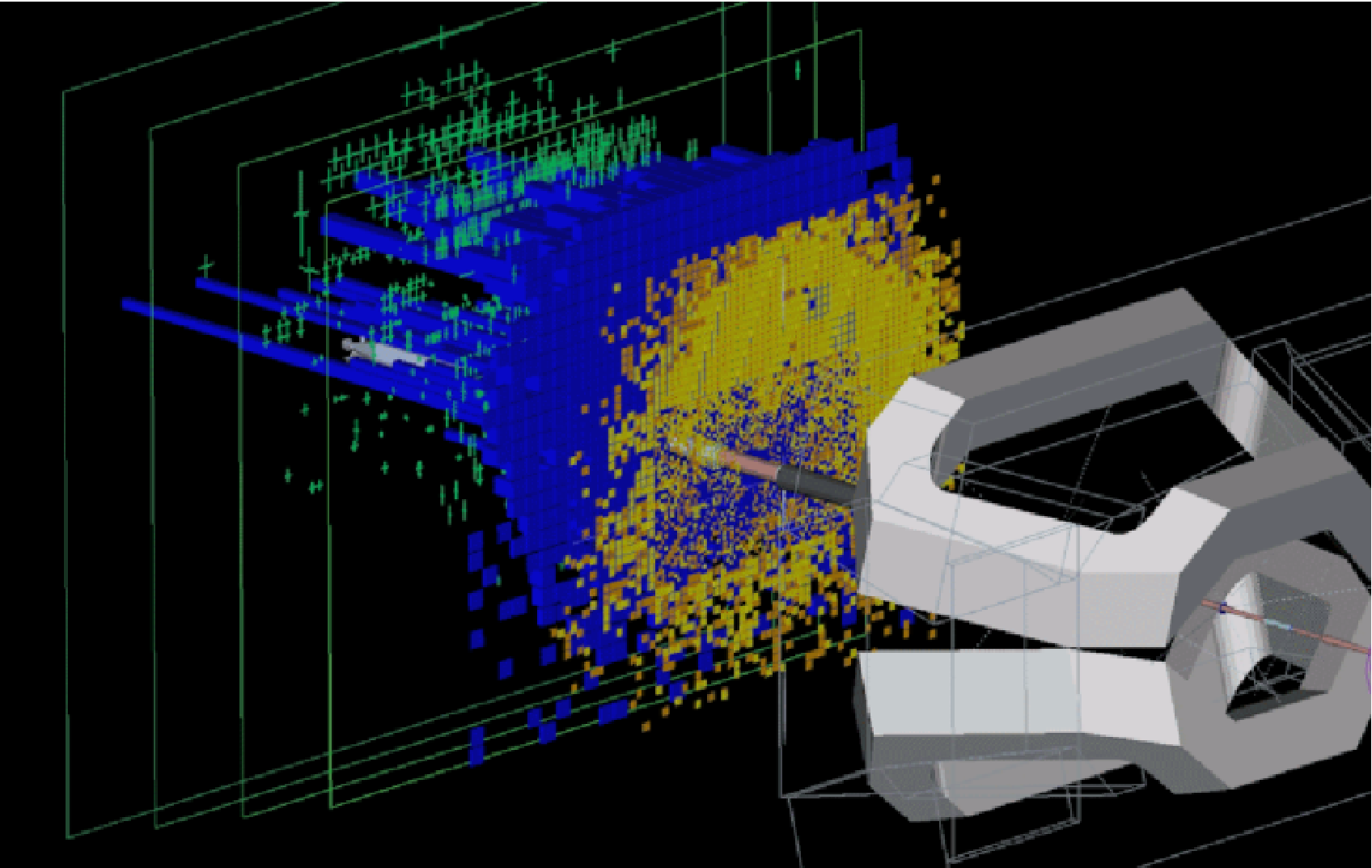,height=40mm}~~~~~~~~
\psfig{figure=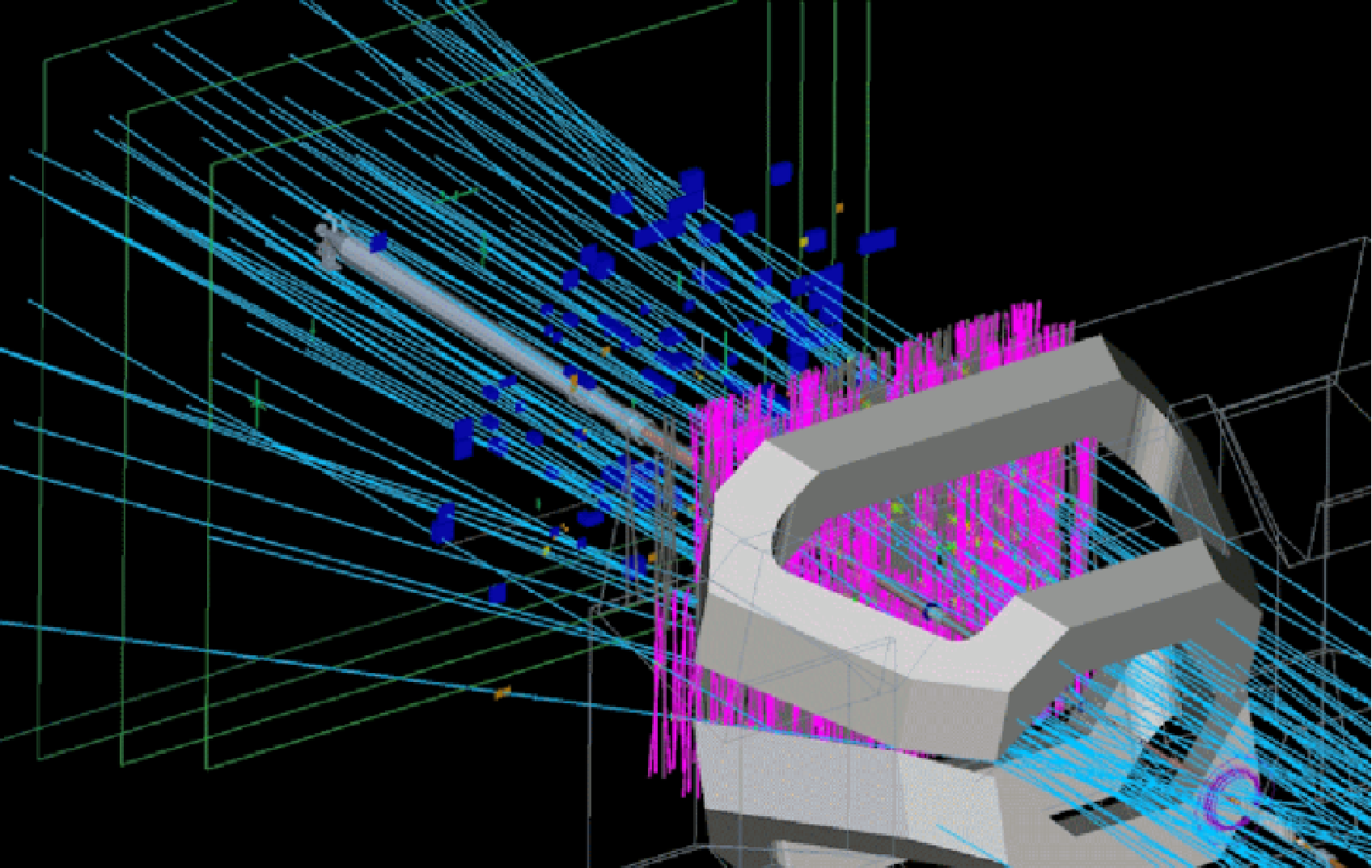,height=40mm}
\caption{Tracks reconstructed by MUON and CALO systems (left), and by the Tracking System (right).}
\label{fig:MediaDay}
\end{center}
\end{figure}

A new machine test was performed on September 10th of 2008: one beam was injected upstream the LHCb experiment
and circulated around the accelerator ring for about 30~minutes. Many clean events have been recorded, and splash
events as well, when the beam was hitting the TDI. Figure~\ref{fig:MediaDay} shows two events recorded by the 
Muon and the Calorimeter systems (left) and by the Tracking System (right).

\section{Conclusions}
The LHCb commissioning started in 2007. In the first phase, each detector subsystem has been individually
commissioned: calibration pulses have been used to test the hardware, the electronic channels and the control 
software as well. In the second phase, the system was exercised as a whole. Despite the horizontal geometry 
of the LHCb detector, we collected more than a million of useful cosmic events that allowed a first time alignment
of most of the subdetectors. Moreover beam induced events during the LHC synchronisation tests provided useful 
data for further time and spatial alignment of the detectors. After a fault was discovered in the LHC, requiring
a shutdown for its repair, LHCb started a phase of general maintenance, looking forward for physics in 2009.

\section*{References}
\begin{thebibliography}{99}
%
\bibitem{magn1} LHCb collaboration, {\it A large Hadron Collider Beauty experiment},
 Technical Proposal, CERN/LHCC 1998-004.
%
\bibitem{ReopTDR} LHCb Collaboration, {\it LHCb Reoptimized Detector Design and Performance}, 
  CERN/LHCC 2003-030.
%
\bibitem{TDRTrigger} LHCb Collaboration, {\it LHCb Trigger System Technical Design Report}, 
CERN/LHCC 2003-031.%
%
\bibitem{L0NIM2} E. Aslanides {\it et al}, \Journal{\NIM}{A 579}{989}{2007}.
%
\bibitem{parkes} C. Parkes {\it et al}, {\it Nucl. Instrum. Methods} {\bf A} (2009).
DOI: 10.1016/j.nima.2009.01.215
%
\end{thebibliography}
\end{document}